\documentstyle[prl,psfig,aps]{revtex}

\begin{document}
% \draft command makes pacs numbers print
\draft

\wideabs{

\title{Exact, numerical, and mean field behavior of a dimerizing 
lattice in one dimension.}
% repeat the \author\address pair as needed
\author{Vasili Perebeinos and Philip B. Allen}
\address{Department of Physics and Astronomy, State University of New York,
Stony Brook, New York 11794-3800}
\author{James Napolitano \cite{byline}}
\address{Commack High School, 10 Radburn Drive, Commack, NY 11725}
\date{\today}
\maketitle
\begin{abstract}

The thermodynamics and dynamics of a one dimensional dimer-forming 
anharmonic model is studied in the classical limit. 
This model mimics the behavior of materials 
with a Peierls instability.
Specific heat, correlation length, and order parameter are calculated 
three ways: (a) by  mean field approximation, (b) by numerical molecular 
dynamics simulation, and (c) by an exact transfer matrix method. 
The single-particle spectrum (velocity-velocity correlation function) is 
found numerically and in mean field theory.

\end{abstract}
}

\section{introduction}

Polyacetylene \cite{Su} and blue bronze \cite{Degiorgi} are examples of 
quasi-one-dimensional materials with broken-symmetry ground states 
\cite{Gruner}, driven by electron-phonon interaction, while 
CuGeO$_3$ \cite{Hase} is driven by a spin-phonon interaction.
In strictly one-dimensional (1D) 
models, phonon fluctuations destroy long range order at temperature $T>0$.
The electron-phonon problem has been extensively studied and 
numerical studies converging towards exact answers have been made 
\cite{Hirsch,Bursill,Gros}. 
Most studies of the electron-phonon problem, especially studies of 
higher dimensional models, omit lattice dynamical fluctuations and 
focus on electronic fluctuation effects, keeping the lattice frozen.
The present paper analyses a model with only lattice-dynamical fluctuations. 
Various authors \cite{Rice,McKenzie} pointed out that 
both electronic and lattice (zero-point and thermal) fluctuations 
should be taken into account to describe the thermodynamics of these 
materials. Lattice fluctuations work as an effective disorder on the 
electronic system, becoming stronger as the $T$ increases 
\cite{Wilkins}.

We take the following Hamiltonian:
\begin{equation}
{\cal H}=\sum_{\ell}\left(-\frac{\kappa_1}{2}x^2_{\ell}+
\frac{\kappa_2}{4}x^4_{\ell}+
\frac{\kappa_3}{2}\left(x_{\ell}+x_{\ell+1}\right)^2+
\frac{p^2_{\ell}}{2M}\right).
\label{hin}
\end{equation}
The first two terms are a single-site (Einstein oscillator) double-well
potential with minima at $\pm u_0$ where $u_0=\sqrt{\kappa_1/\kappa_2}$.
The next term is a second-neighbor spring which prefers displacements
to alternate, $u_{\ell}=(-1)^{\ell} u_0$.  
After introducing dimensionless atom displacements
$\tilde{x}_n=x_n/u_0$,
time $\tilde{t}=t\left(\kappa_1/M\right)^{1/2}$,
and energy $\tilde{E}=E/\kappa_1 u_0^2$, only one free parameter 
$\xi=\kappa_3/\kappa_1$ is left in the problem.
This model was frequently studied in 
the past \cite{Gillis,Scalapino,Koehler,Krumhansl}. 
These studies have been reviewed by Dieterich \cite{Dieterich}. 
McKenzie \cite{McKen} considered a continuous version of the model.

Here we solve this problem numerically by molecular dynamics (MD)
simulation and compare with a mean field 
approximation (MFA) as well as
an exact transfer matrix (TM) method.
The MD simulation enables us to
evaluate a dynamical correlation function 
not available by the TM method;
we compare it with the MFA.

\section{Mean Field Solution.}

Exact thermodynamic calculations of
Hamiltonians like (\ref{hin}) are possible only for a 1D lattice,
and often only in the classical limit.
A variational approach employing the Gibbs-Bogoliubov inequality
\cite{Falk,Gillis} 
can approximate the thermodynamics in all dimensions, either quantum
or classical.
Define a function $\Phi$ which bounds the exact free energy function 
$F(T)$ from above:
\begin{equation}
F(T)\leq \Phi \left(\alpha_i,T \right)=F_0+\left<{\cal H}-
{\cal H}_0\right>_{0},
\label{GBenq}
\end{equation}
where ${\cal H}_0$ is a trial Hamiltonian which depends on 
adjustable parameters $\{a_i\}$ used to minimize the right hand 
side of the expression (\ref{GBenq}).
$F_0$ is the free energy of the trial Hamiltonian.  The average $<>_0$
is taken with respect to ${\cal H}_0$.  If ${\cal H}_0$ is harmonic,
then minimization leads to temperature-dependent frequencies 
of ${\cal H}_0$.  This procedure is also known as the 
``self-consistent phonon method.''  We choose the trial Hamiltonian
\begin{equation}
{\cal H}_0=\sum_{\ell}
\left(\frac{\omega_0^2y_{\ell}^2}{2}+
\frac{\omega_0^2\delta^2}{2}\left(y_{\ell}+y_{\ell+1}\right)^2+
\frac{v_{\ell}^2}{2}\right).
\label{htr}
\end{equation}
All quantities are dimensionless; $y_{\ell}$ is a displaced position
variable $\tilde{x}_{\ell}+(-1)^{\ell}u$.
If one raises $T$, the dimerization amplitude
$u$ goes to zero at a transition temperature $T_c$;
${\cal H}_0$ (Eq. \ref{htr}) describes uncoupled 
oscillators with frequency 
$\omega_k^2=\omega_0^2(1+4\delta^2\cos^2(ka/2))$.
We study here only the classical limit for comparison 
with the MD simulation and the TM solution.
The values of $\omega_0$, $\delta$ and $u$ which minimize 
(\ref{GBenq}) depend on $\xi=\kappa_3/\kappa_1$ and on the 
inverse temperature ${\beta}$.
Let $\Phi_{MF}$ denote the minimum value of the trial free energy 
$\Phi$.   From this we can compute the specific heat in MFA,
$C_{\rm MF}=-T\partial^2\Phi_{MF}/\partial T^2$.
We define a displacement correlation function $G$ as
\begin{equation}
G(m)=\frac{1}{N}\sum_{\ell}\left<(-1)^{\ell}x_{\ell}(-1)^{\ell+m}x_{\ell+m}
\right>.
\label{fcor}
\end{equation}
The square of the order parameter is the correlation 
function at large distance $G\left(m\rightarrow \infty\right)$.  In MFA 
it is the variational parameter $u^2$.
The correlation length, defined as decay rate of the correlation 
function at large $m$, is given in MFA by 
\begin{eqnarray}
\ell_{\rm MF}^{-1} & = & \ln\left(
\frac{1+2\delta^2+\sqrt{1+4\delta^2}}{2\delta^2}\right)
\label{GBcllcor}
\end{eqnarray}
Using the phonon dispersion law of the trial Hamiltonian 
(\ref{htr}) we define a
density of phonon states ${\cal D}_{\rm MF}(\omega)=(a/2\pi)dk/d\omega$, 
which is a function of temperature-dependent parameters 
$\omega_0$ and $\delta$.

Another quantity of the interest is the function
\begin{equation}
F(x)=\frac{1}{N}\left<
\sum_{\ell}\delta\left(x-x_{\ell}+x_{\ell+1}\right)
\right>
\label{nnf}
\end{equation}
which gives the distribution of the separation between nearest atoms.
At the zero temperature 
$x_{\ell}=u(-1)^{\ell}$, and distribution is two delta 
functions $F(x)=\left(\delta(x-2u)+\delta(x+2u)\right)/2$. At non-zero 
temperature it can be evaluated in MFA 
\begin{equation}
F_{\rm MF}(x)= 
\frac{(8\pi)^{-1/2}}{\sqrt{\left<\left(y_{\ell+1}-y_{\ell}\right)^2\right>}}
\exp \left(\frac{-(x\pm2u)^2}
{2\left<\left(y_{\ell+1}-y_{\ell}\right)^2\right>}
\right).
\label{nnfgb}
\end{equation}

\section{Transfer matrix method.}  

The TM method has been applied to one 
dimensional anharmonic systems \cite{Scalapino,Krumhansl}.
To evaluate the classical  partition function, one first writes it as
\begin{eqnarray}
Z &=& \prod_{i=1}^N\int dz_i \exp\left(-\beta f\left(z_{i+1},z_i\right)\right)
\label{Z}
\\
f\left(z_{i+1},z_i\right) &=& -\frac{1}{2}z^2_{i+1}+
\frac{1}{4}z^4_{i+1}+
\frac{\xi}{2}\left(z_{i+1}-z_{i}\right)^2
\label{stat}
\end{eqnarray}
where $z_i$ is $(-1)^i \tilde{x}_i$  and obeys periodic 
boundary conditions $z_{N+1}=z_1$. 
Integration of  $Z\times \delta\left(z^{\prime}-z_1\right)$ 
over the variable $z^{\prime}$ 
simplifies the process when
the delta function is expanded in a complete set of functions
\begin{eqnarray}
\delta\left(z^{\prime}-z_1\right)  &=& \sum_n\Psi^*_n\left(z^{\prime}\right)
\Psi_n\left(z_1\right)
\label{dfun}
\end{eqnarray}
where the functions  $\Psi_n(z)$ satisfy the integral equation
\begin{eqnarray}
\int dz^{\prime} \exp\left(-\beta f\left(z,z^{\prime}\right)\right)
\Psi_n\left(z^{\prime}\right) &=& \exp\left(-\beta\varepsilon_n\right)
\Psi_n\left(z\right)
\label{inteq}
\end{eqnarray}
Then the answer for the partition function 
in the thermodynamic limit $N \rightarrow \infty$ 
is $Z=\exp\left(-N\beta \varepsilon_t\right)$,
where $\varepsilon_t$ is the ``ground state'' 
($\exp\left(-\beta \varepsilon_t\right)$ is the largest eigenvalue) of the 
integral equation (\ref{inteq}).
The leading term of the displacement correlation function is \cite{Scalapino} 
\begin{eqnarray}
G(m) &=& |\left<\Psi_t^{\prime}|y|\Psi_t\right>|^2\exp\left(-ml_{\rm c}\right)
\label{ftrans}
\\
l_{\rm c}&=&\frac{1}{\beta\left(\varepsilon^{\prime}_t-\varepsilon_t\right)}
\label{ltrans}
\end{eqnarray}
where $\Psi_t^{\prime}$ (and $\varepsilon^{\prime}_t$)
are the first excited state of the integral equation 
(\ref{inteq}).  Then $l_{\rm c}$ is the correlation length.

We make the kernel  of the integral equation 
symmetric by transforming the wavefunction:
\begin{eqnarray}
\Upsilon(z)=\Psi(z)\exp\left(\frac{\beta}{2}\left(-\frac{1}{2}z^2+
\frac{1}{4}z^4\right)\right)
\label{newpsi}
\end{eqnarray}
To find the smallest $\varepsilon_n$ we  use a variational trial function:
\begin{equation}
\Upsilon_{\rm tr}(z)=
\exp\left(-\frac{b_1}{4}z^4+\frac{b_2}{2}z^2\right)
\sum_{j=0}^m\left(\alpha_j z^{2j}\right).
\label{trialGS}
\end{equation}
Here we have introduced a set of 
adjustable parameters $b_1$, $b_2$, $\{\alpha_j\}$ used to minimize 
$\varepsilon_{\rm tr}\left(b_1, b_2, \{\alpha_j\}\right)$ defined as:

\begin{eqnarray}
\exp\left(-\beta\varepsilon_{\rm tr}\right)=\frac
{\int\int dz dz^{\prime} \Upsilon_{\rm tr}(z)K(z,z^{\prime})
\Upsilon_{\rm tr}(z^{\prime})} {\int dz \Upsilon_{\rm tr}^2(z)}
\nonumber \\
K(z,z^{\prime})=\exp\left(-\frac{\beta}{2}
\left(f(z,z^{\prime})+f(z^{\prime},z)\right)\right)
\label{epsGS}
\end{eqnarray}

In the decoupled case $\xi=0$ the trial wave function (\ref{trialGS}) is exact 
with $b_1=b_2=\beta/2$ and $m=0$. 
To find the first excited state we use an orthogonal function 
$\Upsilon^{\prime}_{\rm tr}=x\Upsilon_{\rm tr}$.
Then we minimize $\varepsilon^{\prime}_{\rm tr}$ which is  defined similarly to  
$\varepsilon_{\rm tr}$
in expression (\ref{epsGS}), but with $\Upsilon_{\rm tr}$ replaced by 
$\Upsilon^{\prime}_{\rm tr}$. 
First we do minimization of (\ref{epsGS})  
only with respect to the parameters $b_1$ and $b_2$ setting all $\alpha_j$ 
to zero  except $\alpha_0=1$. 
Then we fix the values of $b_1$, $b_2$ and minimize 
equation (\ref{epsGS}) with respect to the set of $\{\alpha_j\}, j=1..m$.
By increasing the  
number of adjustable parameters $m$ we find the converged value of 
$\varepsilon_t={\rm min}\{\varepsilon_{\rm tr}\}$.
Usually $m=10$ is sufficient to get an 
accurate answer with a relative difference 
less than $10^{-10}$ between subsequent steps.

Knowing the free energy  $\varepsilon_t$ we 
can deduce the specific heat by taking its second derivative 
with respect to temperature. Using the fact that $\varepsilon_t$ is an 
extremum with respect to the adjustable parameters, we can evaluate its first
derivative  by computing the partial derivative with respect to  
temperature of the right hand side of   
Eq. (\ref{epsGS}).
Then a second derivative is found from finite differences of the computed 
first derivatives.

\section{ Numerical simulation.} 

We solved Newton's 
equations for the chain of atoms governed by  
Hamiltonian (\ref{hin}).
The system was a chain of 250 atoms with periodic 
boundary conditions.  
A small time step $dt=0.014$ was used. 
Each simulation 
lasted for time 70,000 ($5\times 10^6dt$).  The position and 
velocity of each atom were updated using the Verlet algorithm 
\cite{Verlet} with error proportional
to $dt^4$.  To control the temperature, the
atoms suffered random collisions with ``gas molecules'' of equal
mass.  A collision was modeled by exchanging an atom's velocity
with a gas molecule's velocity, which was randomly generated according to
the Maxwell-Boltzmann distribution.
Collisions occurred randomly but on an average of once every 10 $dt$; the
atom that suffered the collision was also chosen randomly.  Atoms were
initialized to be in the positions
$\left(x_{\ell}=(-1)^{\ell} u_0\right)$,  
with random initial velocities. 
The specific heat was calculated using 
\begin{equation}
C = \frac{<E^2>-<E>^2}{T^2}
\label{cnum}
\end{equation}
Each
simulation was divided into 4 periods of 17,500 time units (1,250,000
$dt$).  Over each of these 4 periods, the total energy E in the system was
measured at 100,000 randomly chosen time intervals.  The specific heat
was calculated for each of the 4 periods.  The 
standard deviation of the 4 calculations served as an error bar.
The correlation function $G(m)$, for $m=0$ to $30$, was calculated 
at 400,000 randomly
chosen time intervals and then averaged.  The value of $-\ln (G(m))$ was
graphed {\it versus} $m$.  The range of $m$ values appropriate for linear 
fitting was chosen by eye.  The slope of the linear
fit was taken as the reciprocal correlation length.

The frequency spectrum was similarly calculated from the velocity
correlation function
\begin{equation} 
\chi(t)=\frac{1}{N}\sum_{i=1}^N \left<v(T+t)v(T)\right>
\label{vvcor}
\end{equation}
It was necessary to calculate $\chi(t)$ from $t=0$ up to 3000
in order for $\chi(t)$ to approach 0.  Each value of $\chi(t)$ 
was calculated 1249
times and then averaged.  At the end of simulation, the frequency spectrum
was calculated by a Fourier transform 

\begin{equation}
{\cal D}(\omega)=\int_0^{\infty} \chi(t) \cos(\omega t) dt
\label{dwnum}
\end{equation}
The nearest neighbor distribution function (\ref{nnf}) was found by averaging 
histograms for the displacement distribution
over 1000 randomly chosen times.

\section{Results and Discussion.}

\begin{figure}
\centerline{
\psfig{figure=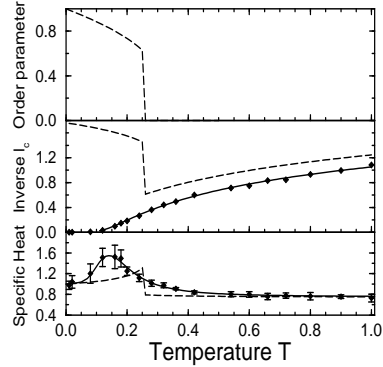,height=2.0in,width=2.0in,angle=0}}
\caption{Specific heat, inverse correlation length and order parameter 
{\it versus} temperature for the case $\xi=\kappa_3/\kappa_1=0.5$. 
The solid lines are exact solutions by the 
TM method; long-dashed line is MFA; 
diamonds are MD results.  TM and MD results agree.}
\label{cv}
\end{figure}

The exact solution has no ordered state at any $T>0$.
In MFA there is a critical 
temperature $T_c$ at which the trial free energies evaluated at 
$u\ne 0$ and $u=0$ 
coincide, indicating a first order transition.  The resulting
order parameter $u^2$ {\sl versus}  
temperature is shown on the top graph of Fig.\ref{cv}.  
The qualitative behavior is similar for all $\xi$, so
we show results only for $\xi=0.5$.

The specific heat in MFA is 
shown on the bottom graph of Fig.\ref{cv} along with TM and 
MD solutions.  The numerical MD and TM specific 
heats agree well.
The uncertainty in the MD specific heat arises from statistical error, 
which should diminish as $N^{-1/2}$, when we take an average over a large 
number $N$ of time slices.

\begin{figure}
\centerline{
\psfig{figure=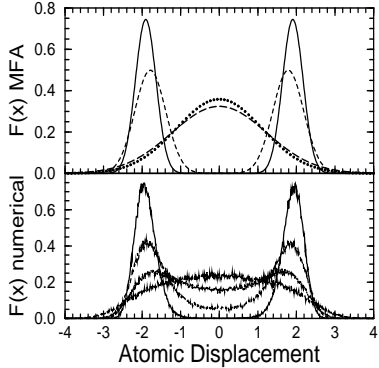,height=2.0in,width=2.0in,angle=0}}
\caption{Nearest neighbor distribution function in
MFA and MD solution for $T$=0.08 (solid line), 
$T$=0.16 (dashed line),
$T$=0.28 (dotted line), $T$=0.8 (long-dashed line).
Atomic displacement is measured in units of  
$\left(\kappa_1/\kappa_2\right)^{1/2}$.}
\label{nn}
\end{figure}

There is a big discrepancy in the exact and MF solutions for specific 
heat, except in high and low $T$ limits, where they coincide. 
To understand why MFA is accurate in the extreme $T$ 
limits, consider the nearest neighbor distribution function 
in MFA and MD cases plotted on Fig. \ref{nn} 
for different $T$. 
At low $T$ atoms are not far from the 
bottom of the well $x_{\ell}=(-1)^{\ell}u_0$. In this situation 
approximation of the actual Hamiltonian (\ref{hin}) by the harmonic version 
(\ref{htr}) works well. In the opposite limit of high $T$ 
harmonic approximation also works well, because the coupling term dominates 
the total energy.  Averages of the 
first and the second terms of the MF Hamiltonian (\ref{htr}), namely 
$\omega_0^2 <y_{\ell}>^2/2$ and 
$\omega_0^2 \delta^2 <y_{\ell}+y_{\ell+1}>^2/2$, coincide 
at $\delta^2=3/4$, which happens at $T=0.74$. 
At $T=0.8$, exact and MFA results agree well for both specific heat and 
nearest neighbor distribution function.
Fig. \ref{nn} shows that the maximum deviation in $F(x)$ happens 
at $T=0.28$, just above $T_c=0.26$. 
The numerical result tells us that 
particles are mostly in the potential wells, while the MFA 
has maximum probability when particles are in 
undisplaced central sites.

Results for the MF, TM, and MD correlation lengths
are plotted on Fig.\ref{cv}. 
The latter two coincide. The error bars of the MD solution come 
from the least square fit of the numerical correlation function.
These error bars are smaller than the size of the 
points and not shown  on Fig.\ref{cv}.
The correlation length diverges at $T=0$ and fits the 1D 
Ising solution 
$A\exp\left(B/T\right)$, 
where coefficients $(A, B)$  depend on  $\xi$, and approximately 
equal (0.18, 0.6) for $\xi=0.5$. 
The MFA gives a  correlation length 
smaller than the exact one at all $T$ and gives wrong behavior at 
$T<T_c$.  Above the MF $T_c$ the predicted $dl_c/dT$ 
is consistent with the exact solution.

The maximum specific heat, which depends on $\xi$, occurs  
in MFA at T$_c$. 
The temperature $T_{\rm max}$ of 
the specific heat maximum in the exact solution is
$\approx 50\%$ smaller than the $T_c$ of MFA, but tracks the MF
transition point.

\begin{figure}
\centerline{
\psfig{figure=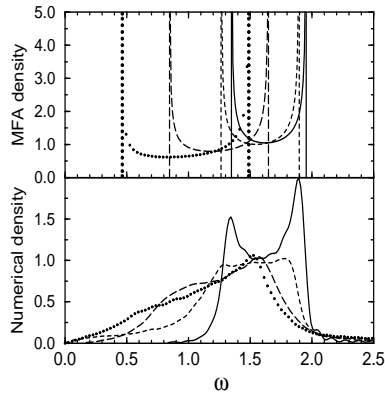,height=2.0in,width=2.0in,angle=0}}
\caption{Density of the phonon states {\it versus} frequency 
(in units of $\left(\kappa_1/M\right)^{1/2}$) in 
MFA and MD for $T$=0.08 (solid line), $T$=0.16 (dashed line),
$T$=0.28 (dotted line), and $T$=0.8 (long-dashed line).}
\label{dw}
\end{figure}

The density of phonon states found by MD and by
MFA are shown on Fig. \ref{dw}. In the low 
$T$ limit again we see agreement between the MFA 
and MD solutions. 
When $T$ is raised, the phonon density in MFA  
becomes broader and moves to lower frequency. 
Since the trial
Hamiltonian (\ref{htr}) does not have translational symmetry, no 
acoustic modes occur; beyond a lower bound, the 
spectrum has no states.  According to MD, a non-zero density of states 
appears at low frequency when $T$ is comparable to the 
double well barrier $T_{\rm B}=0.25$. 
Particles with energy $\geq T_{\rm B}$ take a very long time to overcome 
the barrier, which results in a contribution to the density of states at low 
frequencies. When we raise $T$ well above $T=0.25$, this low 
frequency contribution disappears, and the spectrum becomes more 
similar to the MFA result.

%\section{conclusion}

\acknowledgements
We thank S. Brazovskii for useful suggestions.
This work was supported in part by NSF grant no. DMR-9725037.


\begin{references}

\bibitem[*]{byline}	 Present address, Rensselaer Polytechnic Institute,
			 Troy, NY 12180.

\bibitem{Su}             W. P. Su, J. R. Schrieffer, and A. J. Heeger, Phys. 
                         Rev. Lett. {\bf 42}, 1698 (1979). 

\bibitem{Degiorgi}       L. Degiorgi, G. Gr$\rm \ddot{u}$ner, K. Kim, 
                         R.H. McKenzie, and 
                         P. Wachter, Phys. Rev. B {\bf 49}, 14754 (1994); 
                	 L. Degiorgi, St. Thieme, B. Alavi, 
                         G. Gr$\rm \ddot{u}$ner,
                         R. H. McKenzie, K. Kim, and F. Levy, Phys. Rev. B 
                         {\bf 52}, 5603 (1995).

\bibitem{Gruner}         G. Gr$\rm \ddot{u}$ner, Rev. Mod. Phys. {\bf 60}, 
                         1129 (1988); G. Gr$\rm \ddot{u}$ner, Density Waves 
                         in Solids (Addison-Wesley, Redwood City, 1994).

\bibitem{Hase}           M. Hase, I. Terasaki and K. Uchinokura,
                         Phys. Rev. Lett. {\bf 70}, 3651 (1993).

\bibitem{Hirsch}         E. Fradkin and J. E. Hirsch, 
                         Phys. Rev. B {\bf 27}, 1680 (1983), 

\bibitem{Bursill}        R. J. Bursill, R. H. McKenzie and C. J. Hamer,
                         Phys. Rev. Lett. {\bf 80}, 5607 (1998), 
                         {\bf 83}, 408 (1999),
                         A. W. Sandvik and D. K. Campbell, 
                         {\bf 83}, 195 (1999).
                     
\bibitem{Gros}           R. Werner and C. Gros,
                         Phys. Rev. B. {\bf 57}, 2897 (1998).
                                               
\bibitem{Rice}		 M. J. Rice and S. Strassler, 
                         Sol. State Comm. {\bf 13},  
                         1389 (1973); M. Nakahara and  K. Maki, Phys. Rev. B 
                         {\bf 25}, 7789 (1982).

\bibitem{McKenzie}       R. H. McKenzie, Phys. Rev. B {\bf 52}, 16428 (1995). 

\bibitem{Wilkins}        R. H. McKenzie and J. W. Wilkins, Phys. Rev. Lett. 
                         {\bf 69}, 1085 (1992).

\bibitem{Gillis}         N. S. Gillis, Phys. Rev. B {\bf 11}, 309 (1975);
			 M. Cohen and T. L. Einstein, Phys. Rev. B {\bf 7} 
                         1932 (1973).

\bibitem{Scalapino}      D. J. Scalapino, M. Sears, and R. A. Ferrell, Phys. 
                         Rev. B {\bf 6}, 3409 (1972).        

\bibitem{Koehler}        T. R. Koehler, A. R. Bishop, J. A. Krumhansl, and 
                         J. R. Schrieffer, Solid. State. Comm. 
                         {\bf 17}, 1515 (1975); 
                         T. R. Koehler and N. S. Gillis, Phys. Rev. B {\bf 13},
                         4183 (1976). 

\bibitem{Krumhansl}      J. A. Krumhansl and J. R. Schrieffer, Phys. Rev. B 
                         {\bf 11}, 3535 (1975);
                         S. Aubry, J. Chem. Phys. {\bf 62}, 3217 (1975).

\bibitem{Dieterich}      W. Dieterich, Adv. in Phys. {\bf 25}, 615 (1976).

\bibitem{McKen}          R. H. McKenzie, Phys. Rev. B. 
                         {\bf 51}, 6249 (1995).

\bibitem{Falk}           H. Falk, Am. J. Phys. {\bf 38}, 858 (1970).

\bibitem{Verlet}         L. Verlet, Phys. Rev. {\bf 159}, 98 (1967).



\end{references}
\end{document}